\documentclass[useAMS,usenatbib]{mn2e}
\usepackage{graphicx}
\usepackage{url}
\usepackage{epstopdf}

\newcommand{\apj}{ApJ}

\newcommand{\mnras}{MNRAS}

\newcommand{\aj}{AJ}

\newcommand{\araa}{Annu.Rev.Astron.Astrophysics}

\def\etal {\it et al.}

\title{Completeness I: revisited, reviewed and revived}

\author[Russell Johnston, Lu{\a' i}s Teodoro and Martin Hendry]{Russell Johnston\thanks{e-mail:
r.johnston@physics.gla.ac.uk}, Lu\'{\a i}s Teodoro and Martin Hendry
\\Department of Physics \& Astronomy,
University of Glasgow, Kelvin Building, Glasgow, Scotland, UK, G12 8QQ.}
\begin{document}
\date{\today}
\pagerange{\pageref{firstpage}--\pageref{lastpage}} \pubyear{2007}
\maketitle

\label{firstpage}
\begin{abstract}
We have extended and improved the statistical test recently
developed by Rauzy for assessing the completeness in apparent
magnitude of magnitude-redshift surveys. Our improved test statistic
retains the robust properties -- specifically independence of the
spatial distribution of galaxies within a survey -- of the $T_c$
statistic introduced in Rauzy's seminal paper, but now accounts for
the presence of both a faint and bright apparent magnitude limit. We
demonstrate that a failure to include a bright magnitude limit can
significantly affect the performance of Rauzy's $T_c$ statistic.
Moreover, we have also introduced a new test statistic, $T_v$,
defined in terms of the cumulative distance distribution of galaxies
within a redshift survey. These test statistics represent powerful
tools for identifying and characterising systematic errors in
magnitude-redshift data.

We discuss the advantages of the $T_c$ and $T_v$ statistics over
standard completeness tests, particularly the widely used
$\cal{V}$/$\cal{V}$$_{max}$ test which assumes spatial homogeneity,
and we demonstrate how our $T_v$ statistic can essentially be
regarded as an improved, cumulative $\cal{V}$/$\cal{V}$$_{max}$
test which makes better use of the magnitude completeness
information in a redshift survey. Finally we apply our completeness
test to three major redshift surveys: The Millennium Galaxy
Catalogue (MGC), The Two Degree Field Galaxy Redshift Survey
(2dFGRS), and the Sloan Digital Sky Survey (SDSS). We confirm that
MGC  and SDSS are complete up to the published (faint) apparent
magnitude limit of $m_{b_j}=20.00$ mag. and $m_{r}=17.45$ mag.
respectively, indicating there are no residual systematic effects
within the photometry.  Furthermore, we show that, unless a bright
limit is included for 2dFGRS, the data-set displays significant
incompleteness at magnitudes brighter than the published limit of
$m_{b_{j}}=19.45$ mag.

\end{abstract}
%**************** INTRODUCTION ***************************************************************************
%*************************************************************************************************************
\section{Introduction}

In recent years the statistical analysis of galaxy redshift surveys
has become a thriving industry in cosmology, yielding powerful
constraints on the parameters of both the underlying cosmological
world model and on the luminosity distribution of galaxies as a
function of redshift, environment and morphological type. However,
both tasks are rendered complicated by the presence of observational
selection effects -- due to e.g. detection thresholds in apparent
magnitude, colour, surface brightness or some combination thereof.
Over many years, therefore, a wide range of statistical tools has
been developed to identify, characterise -- and hopefully to remove
-- the impact of observational selection effects from
magnitude-redshift surveys.

Of particular interest are data-sets which are complete in apparent
magnitude -- meaning that all galaxies brighter than some specified
limiting apparent magnitude (or, as is pertinent to this paper, with
apparent magnitudes lying between some specified bright and faint
limiting values) have been observed.  The case of magnitude-redshift
data truncated by a faint apparent magnitude limit has been
extensively discussed in the literature and well-established
techniques have been developed for reconstructing the galaxy
luminosity function in this case.  These include, for example, the
$C-$ method of \cite{1971:LyndenMNRAS.155...95L}, the maximum likelihood fitting
method of \cite{1979:Sandage:ApJ...232..352S} and the stepwise maximum likelihood
method of \cite{Efstathiou:1988_0}.  However, these methods are
formulated under the assumption that the survey data are complete in
apparent magnitude;  hence, the assumption of magnitude completeness
must be rigourously checked.

A classical test for completeness in apparent magnitude is to
analyse the variation in galaxy number counts as a function of the
adopted limiting apparent magnitude (Hubble 1926\nocite{Hubble:1926ApJ....64..321H}). This test, which
presupposes that the galaxy population does not evolve with time and
is homogeneously distributed in space, is however not very
efficient. More specifically, it is difficult to decide in practice
whether deviations from the expected galaxy number count is indeed
an effect of incompleteness in apparent magnitude, or is due to
galaxy clustering and/or evolution of the galaxy luminosity
function, or is indeed created by incompleteness in apparent
magnitude. Of course in designing a completeness test one can also
make use of distance information via galaxy redshifts; the
well-known $\cal{V}$/$\cal{V}$$_{max}$ test of \cite{Schmidt:1968ApJ...151..393S} does this, and
considers -- for a specified magnitude limit -- the ratio of two
volumes: the volume of a sphere of radius equal to the actual
distance of observed galaxy, divided by the volume of a sphere of
radius equal to the {\em maximum\/} distance at which the galaxy
would be observable -- i.e. at the apparent magnitude limit.  It
follows that -- for a non-evolving, homogeneous distribution of
galaxies -- the expected value of $\cal{V}$/$\cal{V}$$_{max}$ is equal to $1/2$.
The $\cal{V}$/$\cal{V}$$_{max}$ test has been used to assess the completeness of
magnitude-redshift samples (see for example \cite{Hudson:1991MNRAS.252..219H},
 but unfortunately it suffers from the same major drawbacks as
the Hubble test based on galaxy number counts: it is difficult to
interpret whether any significant measured departure from the
expected value of $\cal{V}$/$\cal{V}$$_{max}$ is due to incompleteness or to
clustering and evolutionary effects.

In an important (although rarely cited) paper, \cite{Efron:1992ApJ...399..345E}
introduced a powerful new approach to analysing
magnitude-redshift surveys. They proposed a non-parametric test for
the independence of the spatial and luminosity distributions of
galaxies in a magnitude-limited sample, which required no
assumptions concerning the parametric form of both the spatial
distribution and the galaxy luminosity function. They applied this
test to a quasar sample, with an assumed apparent magnitude limit,
in order to robustly estimate the cosmological parameters
characterising the luminosity distance-redshift relation of the
quasars (see also \citep{1998astro.ph..8334E}).

\cite{Rauzy:2001MNRAS.324.51R} noted that the essential ideas of Efron \& Petrosian
could be straightforwardly adapted and extended to turn their
non-parametric test of the cosmological model into a non-parametric
test of the assumption of a magnitude-limited sample -- thus
developing a simple but powerful tool for assessing the magnitude
completeness of magnitude-redshift surveys. As was the case with
Efron \& Petrosian -- and unlike the Hubble number counts or
$\cal{V}$/$\cal{V}$$_{max}$ tests -- the Rauzy test statistic requires no
assumption that the spatial distribution of galaxies is homogenous.
Moreover, it also requires no knowledge of the parametric form of
the galaxy luminosity function. On the other hand, the Rauzy test
was formulated only for the case of a sharp, faint apparent
magnitude limit.  The aim of this paper will be to extend Rauzy's
formalism to account for a bright apparent magnitude limit, and to
apply the extended completeness test to several recent galaxy
redshift surveys.

This paper is, therefore, organised as follows.  In Section 2 we
review the completeness test introduced by \cite{Rauzy:2001MNRAS.324.51R} and extend
it to the case of a galaxy survey with both a faint and a bright
apparent magnitude limit.  In Section 3 we then introduce a further
variant on the original Rauzy completeness test, which is based on
the cumulative distance distribution of observable galaxies in a
magnitude-redshift survey. In Section 4 we briefly describe the
properties of three recent redshift surveys: the Millennium Galaxy
Catalogue, the Sloan Digital Sky Survey and the Two Degree Field
Galaxy Redshift Survey. In Section 5 we then apply our new
completeness test to these three surveys, investigating their
completeness in apparent magnitude. Finally, in Section 6 we
summarise our conclusions.
\begin{figure*}
    \begin{center}    
         \scalebox{0.44}{
     \includegraphics{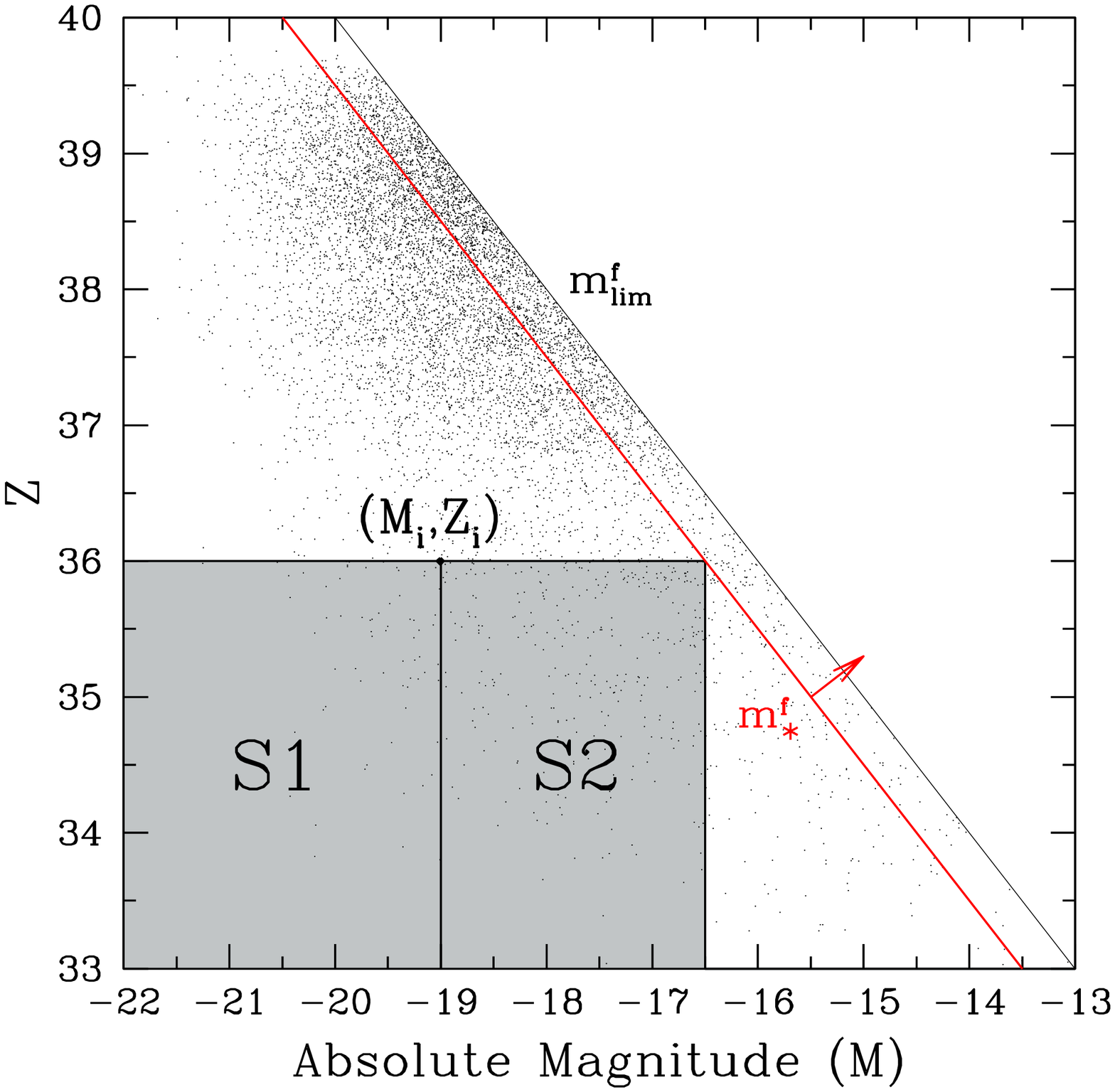}
     \includegraphics{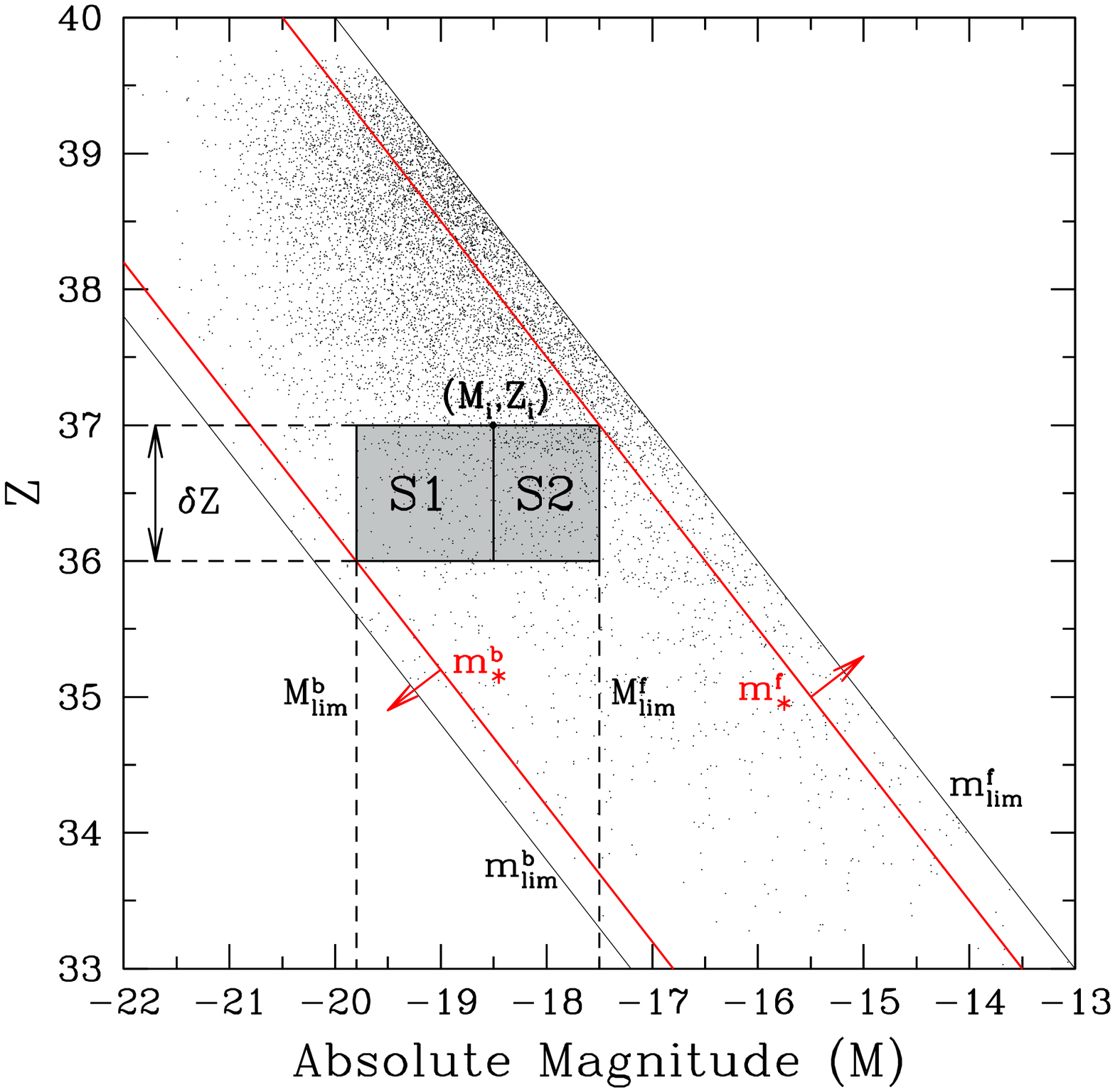}
        }
      \caption{\small Schematic diagram illustrating the construction
       of the rectangular regions $S_1$ and $S_2$, defined for a typical galaxy
        at $(M_i,Z_i)$.  The left panel shows the original R01 construction of regions $S_1$ and $S_2$.  
        The right hand panel shows the construction of regions $S_1$ and $S_2$ with the inclusion
      of a bright limit.  These regions are uniquely defined
      for a slice of fixed width, $\delta Z$, in corrected distance modulus,
      and for `trial' bright and faint apparent magnitude limits $m^{\rm b}_*$
      and $m^{\rm f}_*$ respectively. Also shown are the true bright and faint
      apparent magnitude limits $m^{\rm b}_{\rm lim}$ and $m^{\rm f}_{\rm lim}$,
      within which the rectangular regions  $S_1$ and $S_2$ contain a joint 
      distribution of $M$ and $Z$ that is separable.}
      \label{fig:mz1}
    \end{center}
  \end{figure*}
%

%**************************************************************************************************************
%**************** THE ROBUST TEST OF COMPLETENESS SECTION **********************************
%**************************************************************************************************************
\section{Extending the Rauzy completeness test}
\label{sec:zeta}

In this section we review briefly the robust completeness test introduced by Rauzy (2001; hereafter R01) and applied in \cite{Rauzy:2001MNRAS.328.1016R}, and extend it to include the effect of an
imposed bright apparent magnitude limit to a magnitude-redshift survey. While this extension is straightforward, our approach in this paper will be rather pedagogical in order to benefit those
readers not previously familiar with the formalism previously developed in R01 and Rauzy, Hendry \& D'Mellow (2001).
\subsection{Assumptions and statistical model}
The fundamental assumption of the Rauzy completeness test is that the luminosity function of the galaxy population does not depend on the 3-D redshift space position $\mathbf{z}=(z,l,b)$ of the galaxies. Although this assumption is restrictive, note that it is common to most classical number counts tests of completeness and indeed (when applied in the context of assessing magnitude completeness) the  $\cal{V}$/$\cal{V}$$_{max}$  test. Moreover, the results derived in Appendix 1 of Rauzy, Hendry \& D'Mellow (2001) imply that the completeness test of R01 remains valid in the case of pure density evolution. Note also that the completeness test remains valid for the case of pure luminosity evolution provided that an
evolution correction is applied to account for the (assumed known) functional dependence of the mean luminosity at given redshift.
For clarity, we will defer until a subsequent paper the interesting case where one wishes {\em simultaneously\/} to assess the magnitude completeness {\em and\/} estimate the parameters of a luminosity evolution model. In this paper, where appropriate, we will apply evolutionary and extinction corrections from the literature to the photometry of the galaxy surveys we consider. Specifically,
following the notation of R01, we introduce the corrected distance modulus $Z$, defined as
\begin{equation}\label{Z}
Z = m - M = \mu(z) + k_{\rm cor}(z) + e_{\rm cor}(z)+A_g(l,b),
\end{equation}
where $\mu(z)$ is the (cosmological model-dependent) distance modulus at redshift $z$, $k_{\rm cor}(z)$ and $e_{\rm cor}(z)$ are $k$-corrections and evolutionary corrections respectively and
$A_g(l,b)$ is an extinction correction dependent on galactic co-ordinates $(l,b)$. Note, however, as will be seen in Section 5 below, that the application (or not) of $k$-corrections and evolutionary corrections generally does not have a strong impact on the performance of our completeness test.
Neglecting for the moment any observational selection effects, the joint probability density in position and absolute magnitude for the galaxy population can be written as
\begin{equation}\label{dP_1}
d{P_{zM}} \  {\propto} \  d{P_{z}} \  \times \ {dP_{M}} \ = \
{\rho}(z,l,b)dldbdz \times  f(M)dM,
\end{equation}
where ${\rho}(z,l,b)$ is the 3D redshift space distribution function of the sources along the past light-cone and $f(M)$ is the galaxy luminosity function, defined following e.g. \cite{Binggeli:1988ARA&A..26..509B}.
We now take as our null hypothesis that the selection effects are separable in position and apparent magnitude, and that the observed sample is complete in apparent magnitude for those objects which are
simultaneously brighter than a specified faint apparent magnitude limit, $m^{\rm{ f}}_{\rm{lim}}$, and fainter than a specified bright apparent magnitude limit, $m^{\rm{b}}_{\rm{lim}}$. Under this null
hypothesis the selection function $\psi(m,z,l,b)$ can be written as 
\begin{equation}\label{Selectioneffects}
\psi(m,z,l,b) \equiv \theta(m^{\rm f}_{\rm lim}-m) \times \theta(m -
m^{\rm b}_{\rm lim}) \times \phi(z,l,b),
\end{equation}
where $\theta(x)$ is the Heaviside or `step' function defined as
\begin{equation}
\theta(x) =    \left \{\begin{array}{ll}
1 &\quad {\rm if} \, \, x \geq 0,\\
0&\quad {\rm if} \, \, x < 0,
\end{array}
\right .	
\end{equation}
and $\phi(z,l,b)$ describes the selection effects in angular position and observed redshift. Taking into account this model for the selection effects, the probability density function describing the joint distribution of absolute magnitude $M$ and corrected distance modulus $Z$ for the observable galaxy population may therefore be written as
\begin{equation}\label{dP_2}
dP = \overline{h}(Z) dZ \, f(M) dM \, \theta(m^{\rm f}_{\rm lim}-m)
\theta(m - m^{\rm b}_{\rm lim}),
\end{equation}
where $\overline{h}(Z)$ is the probability density function of $Z$ for observable galaxies, marginalised over direction on the sky,
i.e.
\begin{equation}
\overline{h}(Z) = \int_l \, \int_b \, h(Z,l,b) dl db,
\end{equation}
and the integrand $h(Z,l,b)$ is equal to the (suitably normalised) product of the 3-D redshift space density $\rho(z,l,b)$ and the selection function $\phi(z,l,b)$, re-expressed as a function of $Z$.

Note from Equation (\ref{dP_2}) that the faint and bright apparent magnitude limits introduce a correlation between the variables $M$ and $Z$ for observable galaxies.

\subsection{Defining the random variable $\zeta$}

As in R01, the key element of our extended completeness test is the definition of a random variable, $\zeta$, related to the cumulative luminosity function of the galaxy population. We proceed in a
similar manner to R01, but now with both a bright and faint apparent magnitude limit. To see how we construct $\zeta$ in this more general case consider Fig.~1, which schematically represents an
$M-Z$ plot of corrected distance modulus versus absolute magnitude for the observable population of galaxies.  The left hand panel shows this plot for the case of a faint apparent magnitude limit only, as was considered in R01. The right hand panel is for the more general case which we consider here.  Shown on the right hand panel are solid diagonal lines representing the `true' faint and bright
apparent magnitude limits, $m^{\rm f}_{\rm {lim}}$ and $m^{\rm b}_{\rm {lim}}$ respectively, while the bold diagonal lines represent putative faint and bright magnitude limits, $m^{\rm f}_*$ and $m^{\rm b}_*$ respectively.  The position $(M_i,Z_i)$ of the $i^{\rm{th}}$ galaxy is also indicated on both panels, and the schematic diagrams are superimposed on the actual $M-Z$ distribution for the Millennium Galaxy Catalogue (Driver et al. 2005; see below).

In graphical terms, the essential idea of our completeness test is to identify from the data the faintest value of $m^{\rm f}_*$ and the brightest value of $m^{\rm b}_*$ which together bound a {\em rectangular\/} region of the $M-Z$ plane, within which the joint distribution of $M$ and $Z$ for observable galaxies is separable. If we compare the left and right hand panels of Fig.~1, we can see that the addition of a bright magnitude limit has an important impact on the construction of this separable region: in short, the
region is no longer unique. However, if we fix the width, $\delta Z$, in corrected distance modulus as shown in the right hand panel of Fig.~1, the corresponding separable region {\em is\/} now uniquely defined. Moreover we can then define for the $i^{\rm th}$ galaxy the following absolute magnitudes:
\begin{itemize}
\item $M^{\rm f}_{\rm lim} (Z_i)$, the absolute magnitude of a galaxy, at corrected distance modulus $Z_i$, which would be observed at the true faint apparent magnitude limit $m^{\rm f}_{\rm lim}$, 
\item $M^{\rm b}_{\rm lim} (Z_i-\delta Z)$, the absolute magnitude of a galaxy, at corrected distance modulus $Z_i - \delta Z$, which would be observed at the true bright apparent magnitude limit $m^{\rm b}_{\rm lim}$.
\end{itemize}
These two absolute magnitudes are indicated, for the {\em
putative\/} magnitude limits $m^{\rm f}_*$ and $m^{\rm b}_*$, by the
vertical dashed lines in the right hand panel of Fig.~1.

We now define the random variable $\zeta$ as follows
\begin{equation}\label{zeta}
\zeta = \frac{F(M) - F ( M^{\rm b}_{\rm lim}(Z - \delta Z) )}{F (
M^{\rm f}_{\rm lim}(Z) ) - F ( M^{\rm b}_{\rm lim}(Z - \delta Z) )},
\end{equation}
where $F(M)$ is the Cumulative luminosity function, i.e.
\begin{equation}\label{CLF}
F(M)=\int_{-\infty}^{M} f(x)dx.
\end{equation}

It is straightforward to show from this definition that the random
variable $\zeta$ has a uniform distribution on the interval $[0,1]$,
and furthermore that $\zeta$ and $Z$ are statistically independent.
Thus $\zeta$ shares the same two defining properties as the
corresponding random variable defined in R01. Equation (\ref{zeta})
therefore generalises the definition of $\zeta$ to the case of a
galaxy survey with bright and faint magnitude limits. The relevance
of $\zeta$ as a diagnostic of magnitude completeness will be
demonstrated in the next two sections.

\subsection{Estimating $\zeta$ and computing the $T_c$ statistic}
\label{estzeta}

As was the case in R01, the random variable $\zeta$ has the very
useful property that we can estimate it without any prior knowledge
of the cumulative luminosity function $F(M)$. Given a value of
$\delta Z$, it is clear from Fig.~1 that for each point
$(M_{i},Z_{i})$ in the $M-Z$ plane we can define the regions $S_1$
and $S_2$ as follows:
\begin{itemize}
\item $S_1 =  \{(M,Z): M^{\rm b}_{\rm lim} \leq M \leq
M_i, \, \, Z_i - \delta Z \leq Z \leq Z_i \},$
\end{itemize}
\begin{itemize}
\item $S_2 =  \{(M,Z): M_i < M \le M^{\rm f}_{\rm lim},
\, \, Z_i - \delta Z \leq Z \leq Z_i \}.$
\end{itemize}
In the special case where there is no bright limit 
the regions $S_1$ and $S_2$
are as shown in the left hand panel of Fig.~1.

Clearly the random variables $M$ and $Z$ are now independent within
each sub-sample $S_1$ and $S_2$. Therefore from Equation (\ref{dP_2})
the number of points $r_i$ belonging to $S_1$ satisfies
\begin{equation}\label{r_i} \frac{r_i}{N_{\rm gal}}= \int_{Z_i-\delta Z}^{Z_i} \overline{h}(Z') dZ' \,\times
\,\int_{M^{\rm b}_{\rm lim}}^{M_i} f(M)\,dM,
\end{equation}
where $N_{\rm gal}$ is the total number of galaxies in the sample.
Similarly the number of points $n_i$ in $S_i=S_1 \cup S_2$ satisfies
\begin{equation}\label{n_i} \frac{n_i}{N_{\rm gal}}= \int_{Z_i-\delta Z}^{Z_i} \overline{h}(Z') dZ' \,\times
\,\int_{M^{\rm b}_{\rm lim}}^{M^{\rm f}_{\rm lim}} f(M)\,dM.
\end{equation}
The integrals over absolute magnitude in Equations (\ref{r_i}) and
(\ref{n_i}) may be rewritten as
\begin{equation}
\int_{M^{\rm b}_{\rm lim}}^{M_i} f(M)\,dM \, = \,  F \left ( M_i (Z_i)
\right ) - F \left ( M^{\rm b}_{\rm lim} (Z_i) \right ),
\end{equation}
and
\begin{equation}\label{eq:int2}
\int_{M^{\rm b}_{\rm lim}}^{M^{\rm f}_{\rm lim}} 
f(M)\,dM \, = \,  F \left ( M^{\rm
f}_{\rm lim} (Z_i) \right ) - F \left ( M^{\rm b}_{\rm lim} (Z_i) \right ).
\end{equation}
Thus, given a pair of `trial' magnitude limits $m^{\rm f}_*$ and
$m^{\rm b}_*$, it follows from Equations (\ref{zeta}) and
(\ref{r_i}) to (\ref{eq:int2}) that an estimate of $\zeta$ for the
$i^{\rm th}$ galaxy is simply the ratio of the number of galaxies
belonging to $S_1$ and $S_1 \cup S_2$ respectively (where 
$M^{\rm b}_{*}$ and $M^{\rm f}_{*}$ replace $M^{\rm b}_{\rm lim}$ and
$M^{\rm f}_{\rm lim}$ in the definition of $S_1$ and $S_2$). In fact an 
unbiased estimate of $\zeta$ for the $i^{\rm th}$ galaxy is (c.f. R01)
\begin{equation}\label{zetaestimate}
{\hat \zeta_i} =  \frac{r_i}{n_i+1}.
\end{equation}
This estimator is identical to that defined in R01; the introduction
of a bright magnitude limit has simply changed the definition of the
random variable $\zeta$ itself and the membership criteria of the
two regions $S_1$ and $S_2$. Thus, provided that both $m^{\rm f}_*
\leq m^{\rm f}_{\rm lim}$ and $m^{\rm b}_* \geq m^{\rm b}_{\rm
lim}$, then under our null hypothesis ${\hat \zeta_i}$ will be
uniformly distributed on $[0,1]$ and uncorrelated with $Z_i$,
exactly as was the case in R01. Moreover the expectation value $E_i$
and the variance $V_i$ of the $\hat \zeta_i$ are given respectively
by
\begin{equation}\label{Expectationandvariance_zeta}
E_i = E(\hat \zeta_i)= \frac{1}{2}, \quad V_i = E\left[ \left( \hat
\zeta_i-E_i   \right)^2\right] =  \frac{1}{12} \,
\frac{n_i-1}{n_i+1}.
\end{equation}
Note that $V_i$ tends towards the variance of a continuous uniform
distribution between 0 and 1 when $n_i$ is large.

As in R01, we can, therefore, combine the estimator ${\hat \zeta_i}$
for each observed galaxy into a single statistic, $T_c$, which we
can use to test the magnitude completeness of our sample for adopted
trial magnitude limits $m^{\rm f}_*$ and $m^{\rm b}_*$. $T_c$ is
defined as
\begin{equation}\label{T_c}
T_c =  {\displaystyle
 \sum_{i=1}^{N_{\rm gal}}
\left ( {\hat \zeta_i} - \frac{1}{2} \right ) }\Bigg/ { {\displaystyle
\left (\sum_{i=1}^{N_{\rm gal}} V_i \right )^{\frac{1}{2}} } }.
\end{equation}
If the sample is complete in apparent magnitude, 
for a given pair of trial magnitude
limits, then $T_c$ should be normally distributed with mean zero and
variance unity.  If, on the other hand, the trial faint (bright)
magnitude limit is fainter (brighter) than the true limit, $T_c$
will become systematically negative, due to the systematic departure
of the $\hat \zeta_i$ distribution from uniform on the interval
$[0,1]$.
%*************** THE RANDOM VARIABLE  TAU ***********************************************************
%
\section{$T_v$: A variant of the Rauzy completeness test}\label{sec:tv}
\begin{figure*}
    \begin{center}
	\scalebox{0.44}{
     \includegraphics{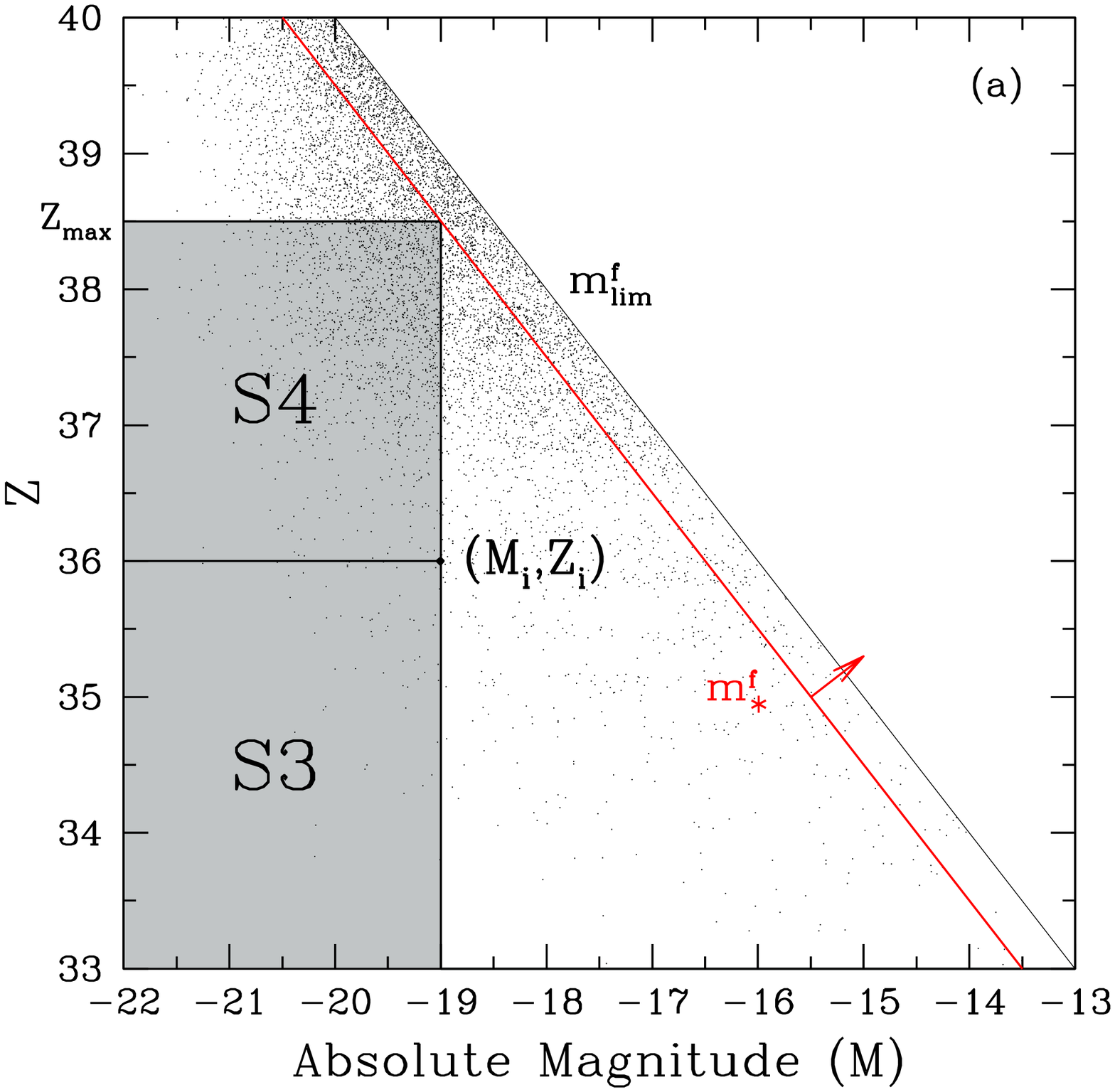}
     \includegraphics{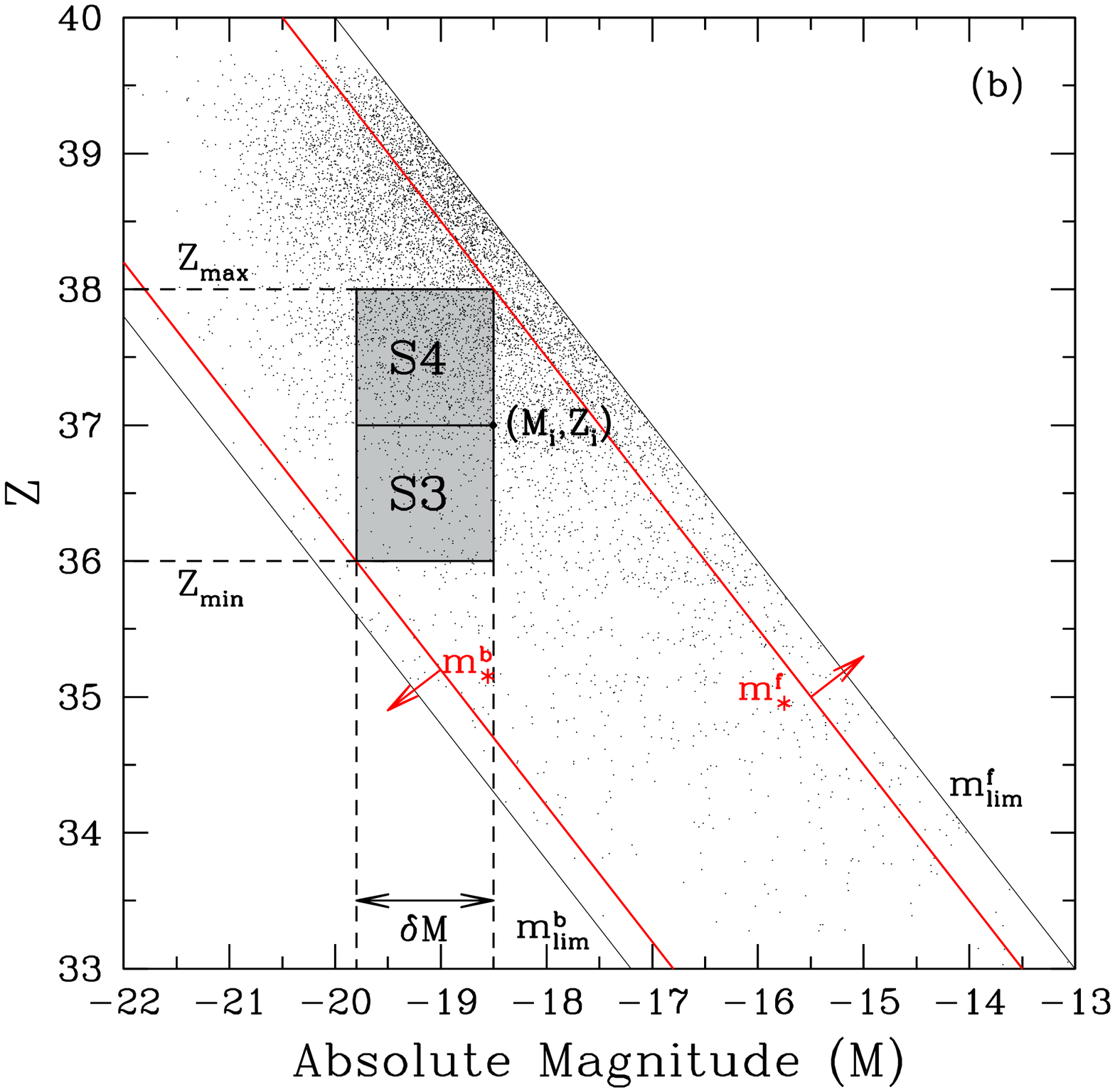}
    	}      
      \caption{\small:  Schematic diagram illustrating the construction
       of the rectangular regions $S_3$ and $S_4$, defined for a typical galaxy
        at $(M_i,Z_i)$, which feature in the estimation of our new 
        completeness test statistic, $T_v$.  Panel (a) illustrates how $S_3$ and 
        $S_4$ are constructed for a survey with only a faint magnitude limit 
	$m^{\rm f}_{\rm lim}$, and are shown for a trial faint limit $m^{\rm f}_\star$.
	Panel (b) illustrates the case where the survey also has a true bright limit	
	$m^{\rm b}_{\rm lim}$, and the rectangles are constructed for trial bright and faint
	limits $m^{\rm b}_{\rm lim}$ and $m^{\rm f}_{\rm lim}$ respectively. Note that the construction of the	 	rectangles is unique for a `slice' of fixed width, $\delta M$, in absolute magnitude.}
      \label{fig:mz2}
    \end{center}
  \end{figure*}
In Section 1 we remarked on the similarities between the $\cal{V}$/$\cal{V}$$_{max}$
 test statistic and the completeness test of R01.  We now
introduce a further variant on the test statistic
$T_c$, related to the distribution of corrected distance modulus for
observable galaxies in a magnitude-redshift survey.

\subsection{Defining the random variable $\tau$}

Fig.~ \ref{fig:mz2} shows schematic $M-Z$ plots, analogous
to Fig.~1: the left panel again has a faint apparent magnitude
limit only while the right hand has a `true' bright and faint
apparent magnitude limit, $m^{\rm b}_{\rm lim}$ and $m^{\rm f}_{\rm
lim}$, with `putative' bright and faint limits, $m^{\rm
b}_{\rm *}$ and $m^{\rm f}_{\rm *}$ respectively, shown as the bold
diagonal lines. Again, the position, $(M_i, Z_i)$, of a typical
galaxy is shown on each panel, and the schematic plots are
superimposed on the actual $M-Z$ distribution of the Millennium
catalogue.

In the right hand panel of Fig.~ \ref{fig:mz2}, however, we
consider a `slice' of width $\delta M$ in absolute magnitude
brighter than $M_i$, as shown. We see that, given $m^{\rm b}_{\rm
lim}$, $m^{\rm f}_{\rm lim}$ and $\delta M$, we can define for the
$i^{\rm th}$ galaxy the following corrected distance moduli:
\begin{itemize}
\item $Z_{\rm upp} (M_i)$, the corrected distance modulus of a galaxy, with
absolute magnitude $M_i$, which would be observed at the true faint
apparent magnitude limit $m^{\rm f}_{\rm lim}$,
\item $Z_{\rm low} (M_i-\delta M)$, the corrected distance modulus of a galaxy,
with absolute magnitude $M_i - \delta M$, which would be observed at
the true bright apparent magnitude limit $m^{\rm b}_{\rm lim}$.
\end{itemize}
These two limiting distance moduli are indicated, for the {\em
putative\/} magnitude limits $m^{\rm f}_*$ and $m^{\rm b}_*$, by the
horizontal dashed lines in Fig.~ 2.

We now define a new random variable $\tau$ as follows. Let $H(Z)$
denote the cumulative distribution function of corrected distance
modulus for observable galaxies, i.e.
\begin{equation}\label{CDF}
 H(Z)=\int_{-\infty}^Z \, \overline{h}(Z') \, dZ'.
\end{equation}
Then $\tau$ is defined as
\begin{equation}\label{tau}
\tau = \frac{H(Z) - H(Z_{\rm low}(M-\delta M))}{H(Z_{\rm upp}(M)) -
H(Z_{\rm low}(M-\delta M))}.
\end{equation}
As was the case for the random variable $\zeta$, it is
straightforward to show that $\tau$ possesses the following
properties:
\begin{itemize}
\item P1: $\tau$ is uniformly distributed between $0$ and $1$,
\item P2: $\tau$ and $M$ are statistically independent.
\end{itemize}
These two properties are exactly analogous to the defining
properties of $\zeta$, except that $\tau$ is now independent of
corrected absolute magnitude, $M$. Once again we can use property P1
to construct a test for completeness in apparent magnitude.
%
%*******************************************************************************************************
\subsection{Estimating $\tau$ and computing the $T_v$ statistic}

Under the assumptions introduced in the previous section, it follows
that $\tau$ can be estimated from our observed data without any
prior knowledge of the spatial distribution of galaxies. To see how
this estimate is constructed, consider again the right hand panel of
Fig.~ \ref{fig:mz2}. For each point with co-ordinates $(M_i,Z_i)$ in
the $M - Z$ plane we can define the regions $S_3$ and $S_4$ as
follows:
\begin{itemize}
\item $S_3 =  \{(M,Z):
M_i - \delta M \le  M \le M_i, \, \,  Z_{\rm low} \le Z \le Z_i \},$
\item $S_4 =  \{(M,Z):
M_i - \delta M \le M \le M_i, \, \, Z_i \le Z \le Z^i_{\rm upp}\}.$
\end{itemize}

In the special case where there is no bright limit 
the regions $S_3$ and $S_4$ are as
shown in the left hand panel of Fig.~1. Indeed, in this case,
$S_3$ is identical to the region $S_1$ shown on the left hand panel
of Fig.~2.

As was the case in Section 2, we see that the random variables $M$
and $Z$ are independent in each sub-sample $S_3$ and $S_4$. Therefore
we can estimate $\tau$ by counting the number, $s_i$, of galaxies
that belong to $S_3$ and the number, $t_i$, of galaxies that belong to
$S_3 \cup S_4$. As in Section 2, an unbiased estimate of $\tau$ is 
given by
\begin{equation}\label{tauestimate}
{\hat \tau_i} =  \frac{r_i}{t_i+1}.
\end{equation}

Thus, provided that both $m^{\rm f}_* \leq m^{\rm f}_{\rm lim}$ and 
$m^{\rm b}_* \geq m^{\rm b}_{\rm lim}$, then under our null hypothesis 
${\hat \tau_i}$ will be
uniformly distributed on $[0,1]$ and uncorrelated with $Z_i$,
exactly as for $\hat{\zeta}_i$. Moreover the 
expectation $E_i$ and variance $V_i$ of the ${\hat \tau_i}$
are respectively
\begin{equation}\label{Expectationandvariance_tau}
E_i = E(\hat{\tau}_i)= \frac{1}{2},
\quad
V_i = E\left[ \left( \hat{\tau}_i-E_i   \right)^2\right] =  \frac{1}{12} \,
\frac{t_i-1}{t_i+1}.
\end{equation}
Again, the variance of $\hat{\tau}_i$ tends 
towards that of a continuous uniform distribution between 0 and 1 for
large $t_i$.

We can, therefore, again combine the estimator ${\hat \tau_i}$
for each observed galaxy into a single statistic, $T_v$, which we
can use to test the magnitude completeness of our sample for adopted
trial magnitude limits $m^{\rm f}_*$ and $m^{\rm b}_*$. $T_c$ is
defined as
\begin{equation}\label{T_v}
T_v =  {\displaystyle
 \sum_{i=1}^{N_{\rm gal}}
\left ( {\hat \tau_i} - \frac{1}{2} \right ) }\Bigg/{ {\displaystyle
\left (\sum_{i=1}^{N_{\rm gal}} V_i \right )^{\frac{1}{2}} } }.
\end{equation}
If the sample is complete in apparent magnitude, 
for a given pair of trial magnitude limits, then 
$T_v$ should be normally distributed with mean zero and
variance unity.  If, on the other hand, the trial faint (bright)
magnitude limit is fainter (brighter) than the true limit, in either
case $T_v$
will become systematically negative, due to the systematic departure
of the $\hat \tau_i$ distribution from uniform on the interval
$[0,1]$.

%*************************************************************************************************************
%******************* THE DATA SECTION ******************************************************************
%*************************************************************************************************************
\section{THE DATA}
We will now apply the tools developed in the previous section to
test the magnitude completeness of three major redshift surveys: 
the Millennium Galaxy Catalogue (MGC), the Two Degree Field 
Galaxy Redshift Survey  (2dFGRS) and The Sloan Digital Sky 
Survey (SDSS-DR1, Early Types).  For ease of comparison we have 
assumed the same background cosmological model but have applied 
existing selection criteria (detailed below) for each survey.
%
%******************* COSMOLOGY SUB SECTION ********************************************************
\subsection{Cosmology}
Unless otherwise stated we have adopted throughout a `Concordance' 
cosmology with present-day matter density $\Omega_{\rm m0} = 0.3$ 
and cosmological constant term $\Omega_{\Lambda0}$ = 0.7, and 
with a value of $H_{0}$ = 100 kms$^{-1}$ Mpc$^{-1}$ for the Hubble 
Constant throughout.  

In order to convert the apparent magnitudes published for each 
survey to corrected absolute magnitudes we apply the following
relation:
\begin{equation}
 M_{i}  = m_{i} - 5\log_{10}(d_{L_{i}})  -  25 - A_g(l,b) - 
k_{\rm cor}(z_{i}) - e_{\rm cor}(z_{i}), 
\end{equation} 
where $d_{L_{i}}$ is the luminosity distance (in Mpc) of the
i$^{th}$ galaxy given by:
\begin{equation}
d_{L_{i}}=(1+z_{i})\left(\frac{c}{H_{0}}\right)\int^{z_{i}}_0
\frac{dz}{\sqrt{(1+z_{i})^{3}\Omega_{m0}+\Omega_{\Lambda0}}},
\end{equation}
and the other terms are as defined in Section 2.1 above.

%******************* SELECTION, K & E CORRECTIONS ***************************************************
\subsection{Selection limits, $k$-corrections and
evolutionary corrections}
\subsubsection{2dFGRS}\label{sec:2dfdata}
The $2dF$ Galaxy Redshift Survey measured redshifts 
using the multifibre spectrograph on the Anglo-Australian 
Telescope. We have used the 2dFGRS public final release 
data-set, from the $`best \  observations'$ spectroscopic 
catalogue, which records redshifts for a total of 245,591 sources. 

The corresponding photometry was taken from the APM galaxy 
catalogue (Maddox et al. 1990\nocite{Maddox:1990}) for galaxies brighter than an 
apparent magnitude of $m_{b_{j}}=19.45$ mag.  The 2dFGRS survey region 
covered two strips: one $75^\circ \times 10^\circ$ around the 
north galactic pole and the other $80^\circ \times 15^\circ$ 
around the southern galactic pole.
\\
\indent
To construct a clean catalogue, we firstly selected those 
galaxies with reliable redshifts -- i.e all galaxies with a 
published redshift quality $Q_{z}\ge3$. We then imposed maximum 
and minimum limits in redshift following Cross et al. 2001 --
i.e. redshifts in the range $0.015<z<0.12$.
From the parent catalogue of 245,591 sources we use a total of 111,082 galaxies.
\\ \indent
There have been a variety of $k$-corrections and evolutionary
corrections applied to the 2dFGRS.  In our analysis we have
adopted those applied by \cite{Cross:2001} and by 
\cite{Norberg:2002a,Norberg:2002b}.
%
%****************************  SLOAN SUB SUBSECTION  **************************************************
\subsubsection{The SDSS early-type galaxy sample}\label{sec:sdss_data}
For our analysis we used galaxies selected from the Sloan Digital Sky 
Survey (SDSS) database.  See Stoughton et al. (2002) \nocite{Stoughton:2002AJ.123.485S} for a 
description of the Early Data Release; 
Abazajian et al. (2003) et al.\nocite{Abazajian1:2003AJ.126.2081A} for a 
description of DR1, the First Data Release; Gunn et al. (1998) \nocite{Gunn:1998AJ.116.3040G}
for a detailed  description of the camera; Fukugita et al. (1996)\nocite{Fukugita:1996AJ.111.1748F}, 
Hogg et al. (2001)\nocite{Hogg:2001AJ.122.2129H} and Smith et al. (2002) \nocite{Smith:2002AJ.123.2121S} for details of the photometric system and calibration; 
Lupton et al. (2001) \nocite{Lupton:2001ASPC.238.269L}  for a discussion of the 
photometric data reduction pipeline; York et al. (2002) \nocite{York:2000AJ.120.1579Y} 
for a technical summary of the SDSS project; Pier et al. (2003)\nocite{Pier:2003AJ.125.1559P}
 for the astrometric calibrations; Blanton et al. (2003)\nocite{Blanton:2003AJ.125.2276B} 
for details of the tilling algorithm; Strauss et al. (2002)\nocite{Strauss:2002AJ.124.1810S} 
and Eisenstein et al. (2001)\nocite{Eisenstein:2001AJ.122.2267E} for details of the target selection. 
\\
\indent
In broad terms, the SDSS sample includes spectroscopic information as well as
photometric measurements in the $u^*, g^*,  r^*, i^*$ and $z^*$ bands. The SDSS First Data
Release covers an area of $\approx$ 2000 deg$^2$ (Abazajian et al.  2003)\nocite{Abazajian1:2003AJ.126.2081A} on the sky.
\\
\indent
The main quantities used in this work are the absolute and apparent magnitudes, and 
redshifts present in the SDSS-First Data Release, early types only (hereinafter referred to as 'SDSS-DR1' ). The selection criteria
has been discussed in \cite{Bernardi:2003AJ.125.1817B} and the data-set compiled in 
\cite{Bernardi:2005AJ....129...61B}. 
39,320 objects have been targeted as early-type galaxies and having dereddened Petrosian  (hereinafter referred to as, $m_{r}$) apparent magnitude (14.5  $< m_r <$  17.75), and  a redshift range of  (0.0 $< z <$ 0.4).  To calculate the distance modulus we assume a Hubble constant of 70 km s$^{-1}$ Mpc$^{-1}$.
%
%******************** MGC SUB - SUBSECTION ***********************************************************
\subsubsection{The Millennium Galaxy Catalogue}\label{sec:MGCDATA}
The Millennium Galaxy Catalogue (MGC) is a medium-deep, $B_j$-band imaging survey, 
spanning 30.9 deg$^2$ that is fully contained within the 2dFGRS and SDSS-DR1.  
The full catalogue contains 10095 galaxies out to a published limiting 
apparent magnitude of  $m_{b_j}=20.00$ mag (e.g. see \cite{Cross:2004} for more detail). 
The photometry was obtained with the Wide 
Field Camera on the 2.5~m Isaac Newton Telescope in La Palma.  The spectroscopy 
was constructed mainly from the redshifts obtained in the 2dFGRS and SDSS.  
In addition, the MGC team 
measured their own redshifts using the spectrograph on the Anglo-Australian 
Telescope for galaxies in the catalogue that had no assigned redshift.
\\ \indent 
Similarly with the 2dFGRS catalogue, only galaxies of a redshift quality $Q_z \ge 3$
have been selected.  For ease of comparison we have imposed the same redshift
cut as \cite{Driver:2005} -- i.e only galaxies in the range $0.013 < z < 0.18$
were included. From the parent catalogue of 10,095 galaxies this yields a
sub set of 7,878 galaxies. Where appropriate we have applied the $k$-corrections
and evolutionary corrections as described in detail by \cite{Driver:2005}.  
%*************************************************************************************************************
%************************** RESULTS***********************************************************************
%*************************************************************************************************************
\section{Results}\label{sec:results}
%
%************************* TESTING THE MGC DATASET**************************************************
\subsection{Testing the MGC dataset}\label{sec:mgc}
\subsubsection{The Rauzy method with a faint limit only}\label{sec:mgc_trad}
\begin{figure}
    \begin{center}
      \includegraphics[scale=0.43]{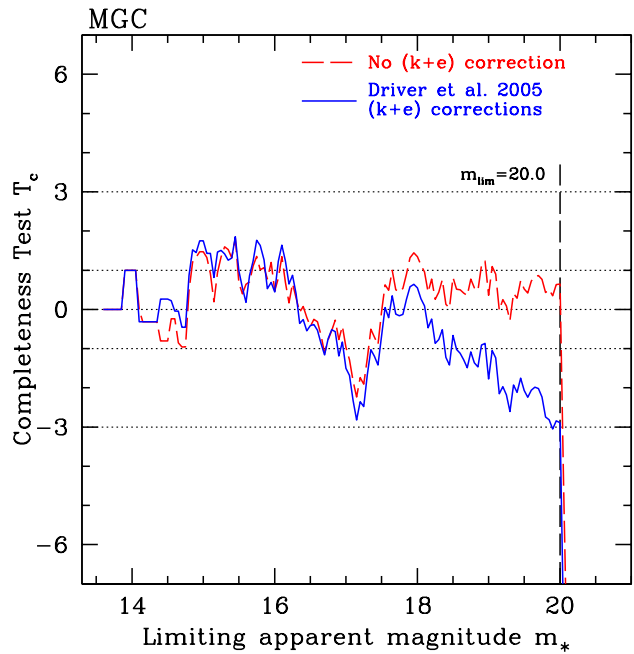}	
      \caption{\small The $T_{c}$ statistic computed for MGC survey.  We have 
               selected galaxies in the range $0.015<z<0.12$ with a quality 
	       $Q_{z} \ge 3$.  The dashed curve is for all galaxies with no $(k+e)$-corrections applied. 
	       The solid curve shows how the $T_{c}$ computed after the application of the $(k+e)$-corrections 	        of  Driver et al. (2005).}
      \label{fig:mgc_tc}
    \end{center}
\end{figure}

Fig.~\ref{fig:mgc_tc} shows the $T_{c}$ statistic as applied to the MGC survey. 
Since there was no bright limit published for this dataset we can use the 
traditional construction of the random variable $\zeta$ as described in R01.  
The dashed curve shows the $T_c$ statistic, as a function of trial magnitude 
limit, computed using apparent magnitudes that have {\em not\/} been 
$(k+e)$-corrected, but have been corrected for 
extinction only.   The figure clearly shows that the $T_c$ statistic remains 
within the $3\sigma$ limits -- consistent with being complete in apparent
magnitude -- up to the published magnitude limit of 20.0 mag., but then drops
very sharply for trial apparent magnitude limits beyond 20.0 mag. 
\\ 
\indent
The solid curve on the same plot again shows the $T_c$ statistic as a function of
trial magnitude limit but now computed for $(k+e)$-corrected apparent magnitudes.  
Although the MGC dataset is still consistent with being complete up to the 
published magnitude limit of 20.0 mag., there is a noticeable departure in the
behaviour of $T_c$ from that for the uncorrected dataset: for trial magnitude 
limits in the range $m_{b_j} \approx 17.5$ to $m_{b_j}=20.00$, $T_c$ 
for the corrected dataset exhibits a slow decline, before again dropping
sharply beyond 20.0 mag.  The most likely explanation for this feature
seems to lie in the way the dataset is selected and corrected.  The raw
dataset, with uncorrected magnitudes, has the same magnitude limit 
imposed on all galaxies independent of their galaxy type.  If, then, 
each galaxy is individually $(k+e)$-corrected, the resultant overall 
magnitude limit for the corrected data will become `fuzzy' without a sharp 
cut-off.  Furthermore, different galaxy populations will be scattered differently, 
leading to a smooth decrease close to the original uncorrected magnitude limit. 
On the other hand, if we do {\em not\/} apply a  $(k+e)$-correction, the 
original magnitude limit remains defined (albeit now without explicitly 
accounting for the effects of evolution and redshifting of each galaxy's
spectral energy distribution). Therefore, to obtain an improved measure of 
completeness which {\em does\/} properly incorporate a $(k+e)$-correction, one
could apply $ROBUST$ separately to subsets of different galaxy type.  This 
would, in principle, lead to the definition of different apparent magnitude
limits for different galaxy types.  In any event, it is clear from 
Fig.~\ref{fig:mgc_tc} that the impact on the inferred `global' apparent
magnitude limit of applying, or not, $(k+e)$-corrections to the MGC dataset is
small.

\begin{figure*}
    \begin{center}
     \scalebox{0.85}{
 	\includegraphics{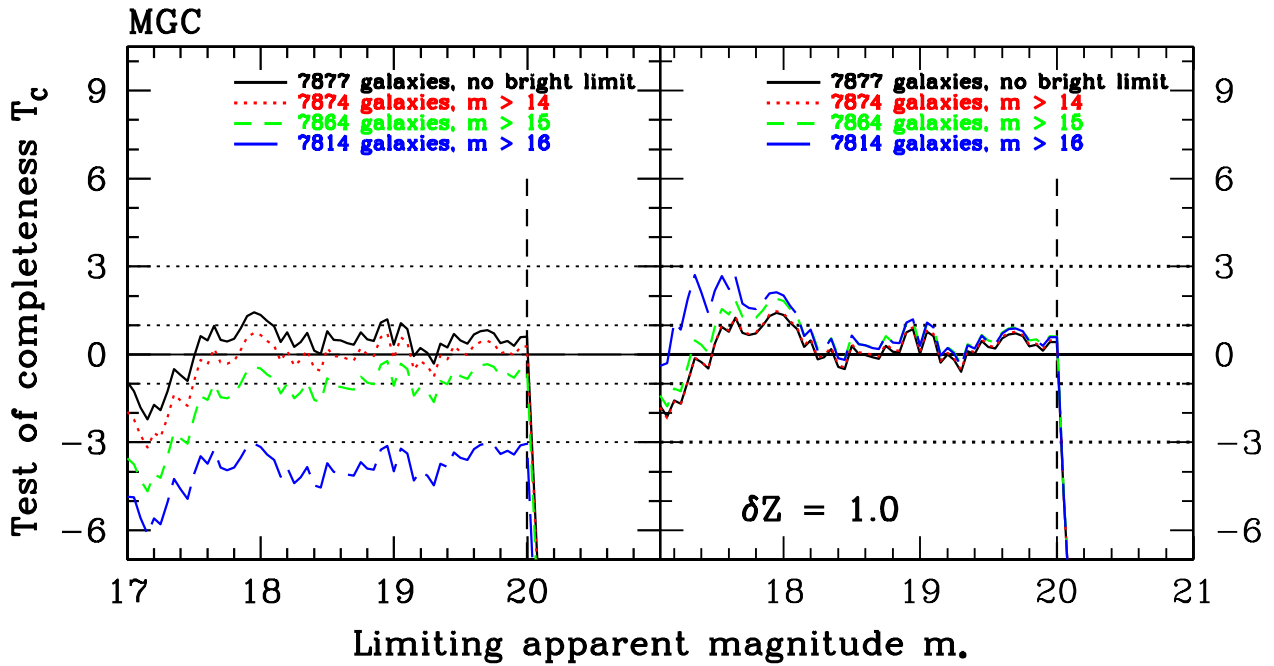} 
	}
      \caption{\small The $T_c$ statistic computed for the MGC survey (without $(k+e)$-corrections) but 
now illustrating the effect, close to the faint limit, of imposing artificially a bright apparent magnitude limit on the selected galaxies. 
In the left panel the solid black curve shows $T_c$ computed assuming no bright limit
(identical to the right hand portion of the dashed curve in Fig. 2) while the other three
curves correspond to progressively fainter bright limits, at $m_{b_j} > 14$, 
$m_{b_j} > 15$ and $m_{b_j} > 16$ respectively.  For all four curves we calculated
$T_c$ following Rauzy (2001) -- i.e. assuming {\em no\/} bright limit.  We can
clearly see that the presence of a bright limit, if ignored, has a significant impact
on the computed value of $T_c$ for faint magnitudes, and thus would adversely affect the
assessment of magnitude completeness close to the faint limit.
In the right hand panel we repeat our analysis for the same four cases as in the
left panel, but now use our extended method which explicitly accounts for the presence
of a bright limit.  We can clearly see that the performance of $T_c$ is no longer
adverseky affected, and a consistent estimate of the faint magnitude limit is obtained
for different imposed bright limits.}
      \label{fig:mgc_nobright}
    \end{center}
\end{figure*}
%
%*******************MGC - BRIGHT LIMIT***************************************************************
%
\subsubsection{Imposing a Bright Limit}\label{sec:mgc_bright}
Having established that MGC is indeed complete up to the published
faint magnitude limit of 20.0 mag., we can now use this survey to
demonstrate how the introduction of a bright limit can affect the
Rauzy completeness test, if not properly accounted for.
\\
\indent To this end, we take the MGC data-set (with no
$(k+e)$-corrections applied) and introduce three, increasingly
faint, artificial bright limits in apparent magnitude: $m_{b_j}>14$,
$m_{b_j}>15$, and $m_{b_j}>16$ respectively.
Fig.~\ref{fig:mgc_nobright}~(left) shows the resulting $T_c$ curves
for the data-sets with these artificial bright limits, but where
$T_c$ was computed assuming {\em no\/} bright limit. The plots
clearly show that, as the bright limit is made progressively
fainter, the computed value of $T_c$ deviates more strongly from the
behaviour expected for a complete data-set. This trend is as
expected: as can be seen in Fig.~1, the presence of the bright limit
breaks the separability of the $M$ and $Z$ distributions for
observable galaxies. Hence, if the bright limit is ignored then the
computed value of $T_c$ will be systematically biased.
\\
\indent We now impose the same artificial bright limits as before
but apply our generalised $T_c$ method which accounts for a bright
{\em and\/} faint limit [see Fig.~\ref{fig:mgc_nobright}~(right)].
It is evident from this plot that, even with a bright magnitude
limit as faint as $m_{b_j}=16$, the performance of the $T_c$
statistic at fainter magnitudes is largely unaffected, showing
consistent behaviour for {\em all\/} the bright limits considered.
%
%**************************** SLOAN DR1**************** ************************************************
%
\subsection{Testing the SDSS-DR1 dataset}
\begin{figure*}
  \begin{center}
    \scalebox{0.85}{
      \includegraphics{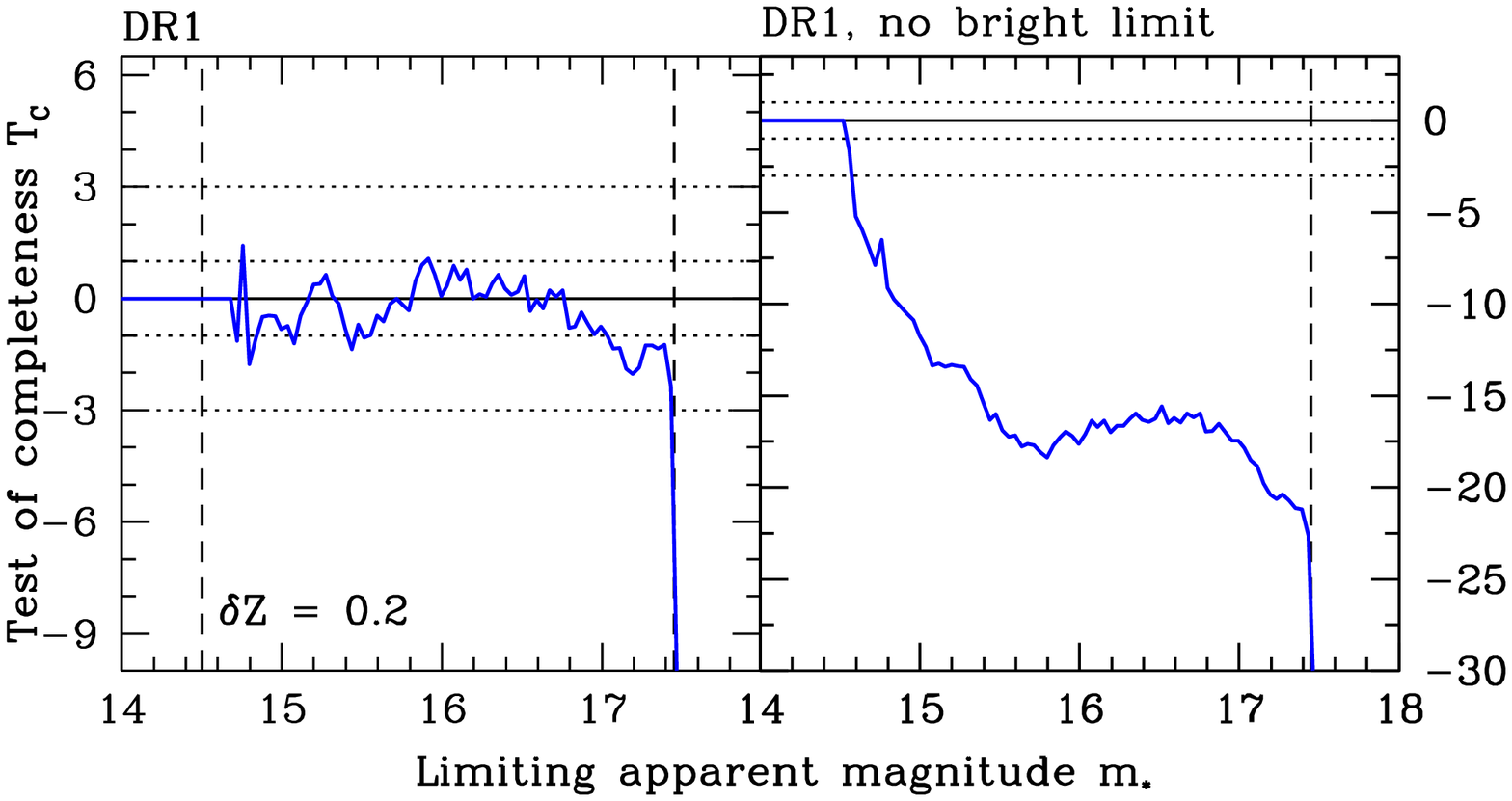}
	 }
    \caption{\small
Performance of the $T_c$ statistic applied, as an illustrative example, to 
the SDSS-DR1 early-type elliptical galaxies.  In the left panel we compute $T_c$
using our extended method, fixing the bright magnitude limit to equal the
published value of $14.55$ mag.  We see that, in this case, the behaviour of $T_c$ 
is consistent with magnitude completeness up to and including the published
faint limit of $17.45$ mag., but the statistic drops rapidly thereafter -- indicating
the sharp onset of magnitude incompletness.  In the right hand panel, on the other
hand, we compute $T_c$ following Rauzy (2001) -- i.e. assuming a faint magnitude limit
only. As anticipated, we see that the test statistic deviates very strongly from
its expected value for a complete data-set at magnitudes which are much brighter
than the published faint magnitude limit (although it is worth noting that $T_c$
still decreases even more rapidly once the published faint limit is exceeded).
}
    \label{fig:sdsstc}	
  \end{center}
\end{figure*}
As previously discussed, the SDSS data-set has both a published
bright and faint apparent magnitude limit of $m_{r}=14.55$ and
$m_{r}=17.45$ mag. respectively. We, therefore, tested the
completeness of the DR1 early type galaxies using our generalised
$T_c$ statistic which accounts for both a faint and bright limit. As
an illustration, we chose to fix the bright limit of the SDSS
data-set to be equal to the published value, and computed the $T_c$
statistic as a function of the trial faint magnitude
limit.  We could, alternatively, have fixed the value of
{\em faint\/} magnitude limit and computed the value of $T_c$ for
different trial values of the bright limit, or indeed we could have
treated simultaneously both the faint and bright limits as free
parameters. However, the case we present is sufficient to illustrate
our method.  Fig.~\ref{fig:sdsstc}~(left) shows the resulting $T_c$
curve. We see that our results are in agreement with the published
faint magnitude limit -- i.e. the behaviour of $T_c$ is consistent
with magnitude completeness up to and including a sharp, faint limit
of $m_{r}=17.45$ mag., followed immediately by the strongly
negative behaviour expected for an incomplete sample at fainter
magnitudes.

The right-hand panel of Fig.~\ref{fig:sdsstc} shows the $T_c$ curve
computed using the traditional Rauzy method with a faint limit only.
Here the results show a similar trend to that seen for the MGC
data-set with an artificially imposed bright limit section, and
further underlines to importance of correctly accounting for a
bright {\em and\/} faint limit when both are present in the data. In
a future paper we will explore in more detail the completeness of
not only the current SDSS data releases, but also how that
completeness varies with bandpass.
%
%******************************  2DF *********************************************************************
%
%
\begin{figure*}
    \begin{center}
 	\scalebox{0.85}{
        \includegraphics{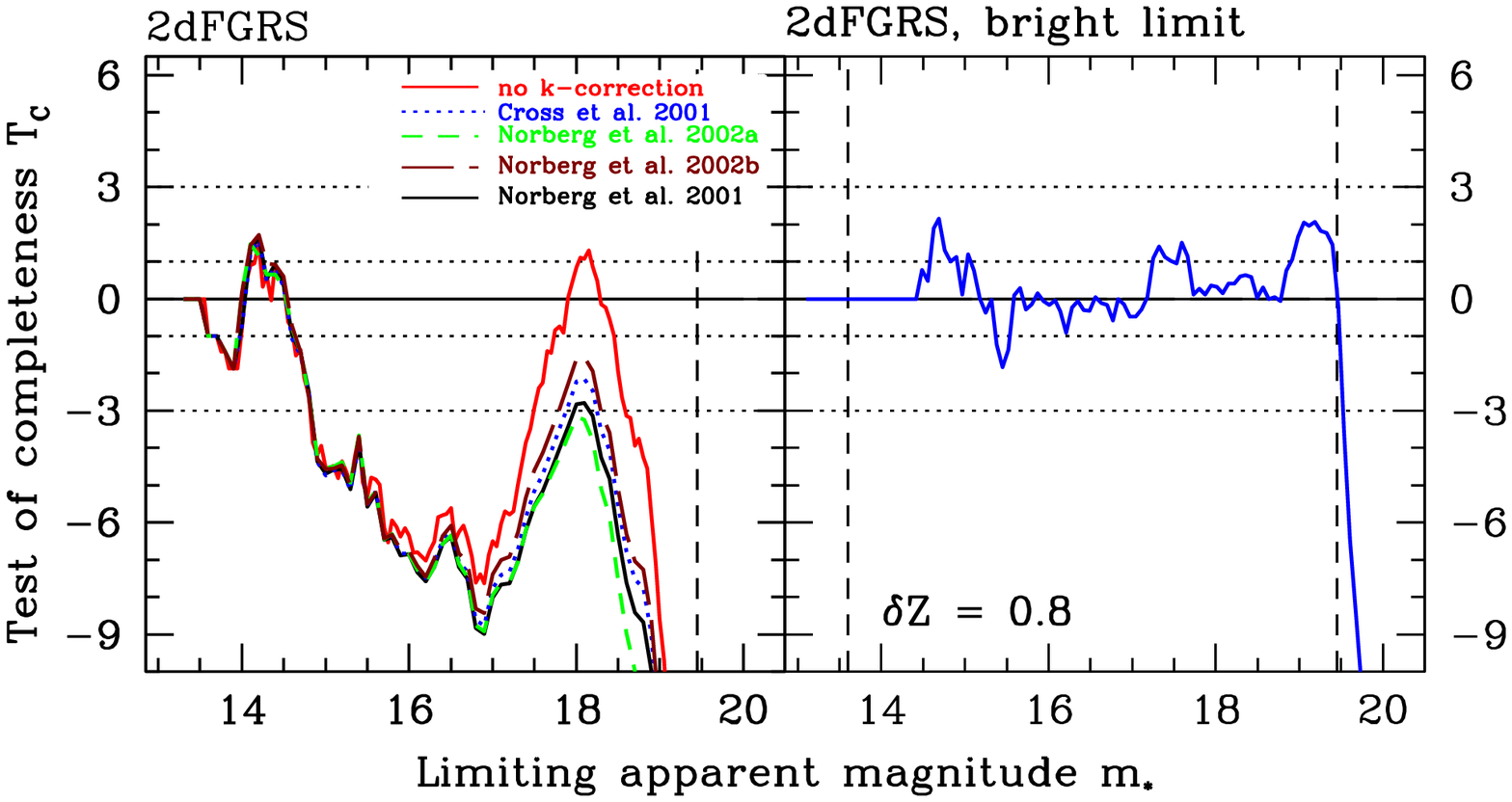}
	 }
        \caption{\small Performance of the $T_c$ statistic applied to our 2dFGRS sample.  In the left
hand panel we compute $T_c$ assuming only a faint magnitude limit, for both
uncorrected magnitudes and after applying various different $(k+e)$-corrections.
Note that several anomalous features are apparent, which are strongly discrepant
from the behaviour of the test statistic expected for a complete sample. Moreover,
$T_c$ begins to drop very sharply around $m \simeq 19.0$ mag. -- significantly
brighter than the published faint magnitude limit of $19.45$ mag.
In the right hand panel we include the effect of a bright apparent magnitude limit,
adopting for simplicity a value equal to the apparent magnitude of the brightest
galaxy in our sample.  The resulting $T_c$ curve (shown for uncorrected magnitudes and computed assuming a fixed `slice' width of $\delta Z = 0.8$ in distance modulus)
is now entirely consistent with magnitude completeness up to and including the
published faint limit, but drops very sharply at fainter magnitudes.}
      \label{fig:2df}   
    \end{center}
  \end{figure*}
\begin{figure*}
    \begin{center}
      \includegraphics[scale=.92]{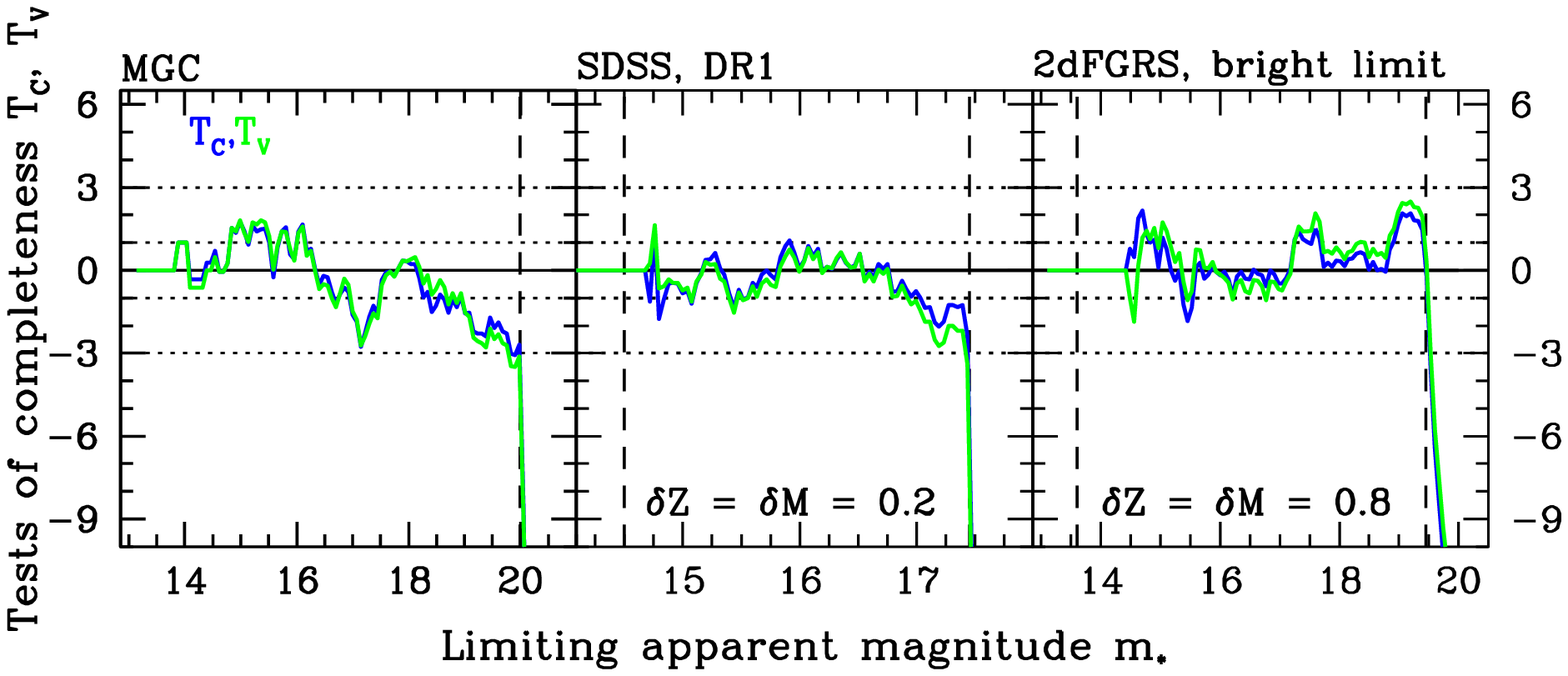}
      \caption{\small Comparison of the $T_c$ and $T_v$ statistics computed for MGC (left panel), SDSS-DR1 
(middle panel) and 2dFGRS (right panel). For the latter two surveys the same bright
limits were adopted as in Figs. 5 and 6 respectively, and appropriate values for
$\delta Z$ (for $T_c$) and $\delta M$ (for $T_v$) were chosen.  Note the almost
identical agreement of the test statistics in each case. To illustrate the robustness
of this result, the MGC results are with $(k+e)$-corrections applied, the SDSS-DR1
results are with $K$-corrections only applied, while the 2dFGRS results are for
uncorrected galaxy data.}
       \label{fig:mgc_tv}
      \end{center}
  \end{figure*}
\subsection{The 2dFGRS Survey}
We move finally to the 2dFGRS survey which, as discussed in the Data
section, has a published faint limit of $m^f_{lim}=19.45$ mag.

\subsubsection{The Rauzy method with a faint limit only}
Our initial approach was to apply the traditional $T_c$ statistic to
the 2dFGRS since the published literature on the survey gives no
indication about the presence of a secondary bright limit.
Fig.~\ref{fig:2df}~(left) shows the behaviour of $T_c$ as a function
of trial magnitude limit, $m^{\rm f}_\star$, for five different
cases. The solid red curve represents the completeness test with
with no $k$- or $e$-corrections applied. The remaining four curves
show $T_c$ with various $(k+e)$- corrections applied to the 2dFGRS
data-set, as shown in the figure key.

Consider firstly the uncorrected data-set (solid red curve). We see
that for $m^{\rm f}_\star < 14.85$ mag. the $T_c$ statistic appears
to behave as we would expect for a complete sample (although of
course this is at the cost of `throwing away' most of the galaxies
in the survey by imposing such a low value for the faint limit).
However, for higher values of $m^{\rm f}_\star$ the statistic drops
dramatically to a minimum value of nearly $8\sigma$ below its
expectation value of zero at $m^{\rm f}_\star = 16.90$ mag. As we
continue to increase $m^{\rm f}_\star$, $T_c$ rises sharply to reach
a peak at $m^{\rm f}_\star = 18.15$ mag, beyond which the statistic
drops dramatically again, exceeding $3\sigma$ below its expected
value at $m^{\rm f}_\star = 18.60$ -- i.e. significantly brighter
than the published magnitude limit of $m^{\rm f}_{\rm lim} = 19.45$
mag.

Could the behaviour of $T_c$ be related to the fact that we have
used an uncorrected data-set? To address this question consider now
the remaining four curves; the dotted and short dashed curves
correspond to the \cite{Cross:2001} and \cite
{Norberg:2001MNRAS.328...64N} global $(k+e)$-corrections
respectively, whereas the long dashed and solid black curves
correspond to the type-dependent $(k+e)$-corrections of  \cite
{Norberg:2002a, Norberg:2002b}. It is clear that the adoption of any
of these corrections has very little effect on the completeness
statistic compared with the uncorrected case. Indeed, if anything,
the addition of $(k+e)$-corrections appears to yield a $T_c$
statistic which is more strongly inconsistent with a complete
sample. This latter effect may be explicable in the same manner as
for the corrected MGC magnitudes described in section \ref{sec:mgc},
although it should be noted that the type-dependent
$(k+e)$-corrections do not appear to perform significantly better
than their global counterparts.

The fact that the value of $T_c$ differs from zero at many standard
deviations over a wide range of trial faint magnitude limits is
clear evidence that the distribution of $M$ and $Z$ for observable
galaxies is {\em not\/} separable with these faint limits. The
physical cause for this is not immediately clear; however, having
demonstrated in the previous subsection the adverse impact on $T_c$
of not correctly accounting for a bright magnitude limit, we next
apply our generalised test statistic to the 2dFGRS data-set.

\subsubsection{2dFGRS with a bright limit}
In the absence of a clear indication from the literature of what is
an appropriate bright magnitude limit, we adopted the brightest
galaxy in our subset (see \S \ref{sec:2dfdata}), $m^{\rm b}_{\rm
lim}=13.60$ mag. The right hand plot of Fig.~\ref {fig:2df} shows
the $T_c$ curve obtained for the 2dFGRS (with no $k$- or $e$- corrections applied)
as a function of trial faint
limit, $m^{\rm f}_\star$, with $m^{\rm b}_{\rm lim}=13.60$ mag. We
have used a $\delta Z$ = 0.8 (a small $\delta Z$ leads to low
numbers of galaxies within the subsets, $S1$ and $S2$, making our
test statistic prone to large statistical fluctuations and therefore
less sensitive to a sharp cut in magnitude). The plot demonstrates a
dramatic change in behaviour of the 2dF completeness, compared with
the traditional $T_c$ statistic. By simply accounting for a bright
limit -- notwithstanding the fact that no published bright limit has
been reported in the literature -- we find that the 2dFGRS data-set
is indeed complete to the published faint magnitude limit with no
evidence for residual systematics.
%
%**************************************************************************************************************
%*************************  APPLICATION OF TV TEST *****************************************************
%**************************************************************************************************************
\section{APPLICATION OF THE $T_v$ STATISTIC}
In the previous sections we introduced and applied our improved
$T_c$ statistic, which accounts for both a faint and bright
magnitude limit in assessing the completeness of a
magnitude-redshift survey. In this section we apply the $T_v$
statistic, introduced in Section 3 above, to the same data-sets. Our
$T_v$ statistic can be thought of as an improved, differential,
version of the classical $\cal{V}$/$\cal{V}$$_{max}$ test of galaxy
evolution, which is generally presented in the literature as
yielding a single number -- the mean value of
$\cal{V}$/$\cal{V}$$_{max}$ averaged over all galaxies in the
survey, adopting a given faint apparent magnitude limit and assuming
that the underlying spatial distribution of galaxies is homogeneous.
In contrast, we can compute $T_v$ as a function of an incrementally
increasing $m^{\rm f}_{\star}$ (and/or indeed, if we wished, an
incrementally {\em decreasing\/} bright limit, $m^{\rm b}_{\rm
lim}$) and thus analyse our data-set via a series of progressively
truncated subsets. Crucially, moreover, like the $T_c$ test and
unlike the $\cal{V}$/$\cal{V}$$_{max}$ statistic, $T_v$ has the
important property of being independent of the spatial distribution
of galaxies within the survey.

Fig.~\ref{fig:mgc_tv} shows a comparison of the $T_v$ and $T_c$
curves for all three surveys.   The left hand plot is the MGC survey
with $(k+e) $-corrections applied.  The $T_v$ curve shows an almost
identical match to the $T_c$ statistic.  Similar behaviour is
evident with SDSS-DR1 and 2dFGRS (middle and right plots).  That
$T_v$ and $T_c$ give a consistent indication of the completeness of
these surveys should not be too surprising, since we are confident
(at least once a bright limit is included in our analysis of 2dFGRS)
that all three are well calibrated and well understood.  Moreover,
they are all relatively shallow in redshift range, which means that
extinction and evolution corrections are not likely to impact too
strongly on our assessment of their completeness (a fact which is
supported by our results for $T_c$). However, one can ask under what
conditions might the two statistics $T_c$ and $T_v$ diverge from
each other?

Consider a galaxy, $i$, characterised by its `coordinates' $(M_i,
Z_i)$. We have two complementary ways of generating volume limited
data-sets for such a pair:
\begin{itemize}
\item at fixed luminosity we can ask what redshift distribution will
produce apparent magnitudes permitted by our selection criteria?
\item alternatively, at fixed redshift we can ask what
distribution of luminosities (i.e. what part of the underlying
galaxy luminosity function) are we sampling, given our selection
limits in apparent magnitude?
\end{itemize}
The former criterion resembles the procedure used to contruct the
$T_v$ statistic, while the latter criterion is more closely related
to the procedure used to construct $T_c$. This also implies that one
might expect the two statistics to behave differently when evolution
becomes important -- simply because evolution will, of course, break
the separability of the {\em underlying\/} joint distribution of $M$
and $Z$, i.e. the conditional distributions of $M$ at given $Z$ and
$Z$ at given $M$ will no longer be simply equal to their marginal
distributions. It seems likely, therefore, that an exploration of
the systematic differences between $T_c$ and $T_v$ for deeper
surveys may be an effective probe of evolution.
%*********************************************************************************************************
%******************** SUMMARY************************************************************************
%*********************************************************************************************************
\section{SUMMARY}
We have developed the completeness test statistic, $T_c$, first
introduced in Rauzy (2001), technique to account for the presence of
both a faint and bright apparent magnitude limit in
magnitude-redshift samples. We have applied it to the MGC, SDSS-DR1
and 2dFGRS surveys. Our results confirm the completeness of
data-sets such as SDSS-DR1 (early types only) where a faint and
bright limit is well defined and published in the literature.
Specifically, we have demonstrated that SDSS-DR1 is complete in
apparent magnitude up to the published magnitude limit of
$m_{r}=17.45$ mag indicating no residual systematics. Similarly, the
magnitude completeness of the MGC survey has also been confirmed up
to its published limit of $m_{b_j}=20.0$ mag. Interestingly,
however, we found that when we incorporated $(k+e)$-corrections to
the MGC data-set, a noticeable (although not statistically
significant) departure from the expected value of the $T_c$
statistic -- and indeed from the computed value of $T_c$ for the
uncorrected data -- was observed close to the magnitude limit. A
possible cause for this effect could be the mixing of galaxy
populations to which a global $(k+e)$ is then applied -- resulting
in a slightly $blurred$ magnitude limit.

Our initial approach to the 2dF galaxy survey was to apply the
original Rauzy test which accounts for a single, faint magnitude
limit only.  This was motivated by the current literature, in which
$only$ a faint limiting magnitude of $m=19.45$ was well defined in
the survey.   However, our the application of our $T_c$ test reveal
that the 2dFGRS is strongly inconsistent with being complete in
apparent magnitude  {\em unless\/} a secondary bright limit
($m=13.6$ in our subset) is included.

Finally, we have also developed another variant on the original
Rauzy $T_c$ completeness statistic, which we denote by $T_v$, based
on the cumulative distance distribution of galaxies in a
magnitude-redshift survey. We find that $T_v$ has potential
advantages over the widely used $\cal{V}$/$\cal{V}$$_{max}$ test:
not least, the $T_v$ statistic retains the same properties as that
of $T_c$ -- i.e. is independent of the spatial distribution of
galaxies within the survey. Furthermore, we have shown by example,
that $T_v$ -- when applied to the same well calibrated and
relatively shallow survey samples as $T_c$ -- produces almost
identical results to that of the $T_c$ statistic. Our future work in
this area will explore the application of $T_v$ and $T_c$ to deeper
surveys, where evolution and $k$-corrections become more important
issues, to investigate the potential of these two statistics as
diagnostics of luminosity and density evolution.
\section*{Acknowledgements}
LT is a Leverhulme Trust Research Fellow at the University of Glasgow.
He would like to thank Vincent Eke and Peder Norberg 
for their infinite patience and enlightening discussions concerning 
the idiosyncrasies of the 2dFRGS.  MAH would like to thank Stephane
Rauzy, Kenton D'Mellow and David Valls-Gabaud for many useful
discussions during the early development and application of the
Rauzy test statistic. RJ would like to thank Simon Driver for providing us with the MGC data-set previous to publication.
\\
\\
 The SDSS-DR1 data-set was kindly provided by Mariangela Bernardi and Ravi Sheth.  Funding for the creation and distribution of the SDSS Archive has been provided by the Alfred P. Sloan Foundation, the Participating Institutions, the National Aeronautics and Space Administration, the National Science Foundation, the U.S. Department of Energy, the Japanese Monbukagakusho, and the Max Planck Society. The SDSS Web site is \url{http://www.sdss.org}. The SDSS is managed by the Astrophysical Research Consortium (ARC) for the Participating Institutions. The Participating Institutions are the University of Chicago, Fermilab, the Institute for Advanced Study, the Japan Participation Group, the Johns Hopkins University, Los Alamos National Laboratory, the Max Planck Institute for Astronomy (MPIA), the Max Planck Institute for Astrophysics (MPA), New Mexico State University, the University of Pittsburgh, Princeton University, the United States Naval Observatory, and the University of Washington.
%**************************************************************************************
\markright{ }
%\bibliography{mnrasmnemonic,bibliography}

\bibliographystyle{astron}

\label{lastpage}
\end{document}